\documentclass[12pt]{article}
\usepackage{amsmath}
\usepackage{graphicx,psfrag,epsf}
\usepackage{enumerate}
\usepackage{natbib}
\usepackage{url} 
\usepackage{color}
\usepackage{bm} 

\newcommand{\blind}{0}

\newcommand{\sfsub}[1]{\scriptstyle{{\sf #1}}} 

\newcommand{\sect}[1]{Section~\ref{#1}}
\newcommand{\fig}[1]{Figure~\ref{#1}}

\newcommand{\curr}{{\sf curr}}
\newcommand{\buff}{{\sf buff}}
\newcommand{\past}{{\sf past}}

\newcommand{\insitu}{{\em in situ\/}}

\begin{document}

\def\spacingset#1{\renewcommand{\baselinestretch}%
{#1}\small\normalsize} \spacingset{1}


\if0\blind
{
  \title{\bf Partitioning a Large Simulation as It Runs}
  \author{Kary Myers\\
    Statistical Sciences, Los Alamos National Laboratory\\
    and \\
    Earl Lawrence \\
    Statistical Sciences, Los Alamos National Laboratory\\
    and \\
    Michael Fugate \\
    Statistical Sciences, Los Alamos National Laboratory \\
    and \\
    Claire McKay Bowen \\
    Applied and Computational Mathematics and Statistics, University of Notre Dame \\
    and \\
    Lawrence Ticknor \\
    Statistical Sciences, Los Alamos National Laboratory \\
    and \\
    Jon Woodring \\
    Applied Computer Science, Los Alamos National Laboratory \\
    and \\
    Joanne Wendelberger \\
    Statistical Sciences, Los Alamos National Laboratory \\
    and \\
    Jim Ahrens \\
    Applied Computer Science, Los Alamos National Laboratory}
    \date{}
  \maketitle
} \fi

\if1\blind
{
  \bigskip
  \bigskip
  \bigskip
  \begin{center}
    {\LARGE\bf Partitioning a Large Simulation as It Runs}
\end{center}
  \medskip
} \fi

\bigskip
\begin{abstract}
As computer simulations continue to grow in size and complexity, they present a particularly challenging class of big data problems.  Many application areas are moving toward {\em exascale} computing systems, systems that perform $10^{18}$ FLOPS (FLoating-point Operations Per Second) --- a billion billion calculations per second.  Simulations at this scale can generate output that exceeds both the storage capacity and the bandwidth available for transfer to storage, making post-processing and analysis challenging. One approach is to embed some analyses in the simulation while the simulation is running --- a strategy often called {\em in situ} analysis --- to reduce the need for transfer to storage. Another strategy is to save only a reduced set of time steps rather than the full simulation. Typically the selected time steps are evenly spaced, where the spacing can be defined by the budget for storage and transfer. This paper combines both of these ideas to introduce an online \insitu\/ method for identifying a reduced set of time steps of the simulation to save. Our approach significantly reduces the data transfer and storage requirements, and it provides improved fidelity to the simulation to facilitate post-processing and reconstruction. We illustrate the method using a computer simulation that supported NASA's 2009 Lunar Crater Observation and Sensing Satellite mission.
\end{abstract}

\noindent%
{\it Keywords:}  complex computer models, piecewise linear fitting, exascale computing, streaming data, online methods, change point detection
\vfill
\hfill {\tiny technometrics tex template (do not remove)}

\newpage
\spacingset{1.45} 

\section{Introduction}
\label{sec:intro}

Scientists routinely build simulations to study complex phenomena. These computer simulations have become increasingly large in scale, keeping pace with increased computing power. As computer simulations grow in size and complexity toward {\em exascale} supercomputing, which will provide $10^{18}$ FLOPS, FLoating-point Operations Per Second (a billion billion calculations each second), our ability to analyze the output of these simulations is frustrated by I/O bottlenecks: both limited disk space and limited transfer bandwidth (e.g., \citet{oldfield2014}). In other words, we don't have the storage capacity to save the full simulation, and we don't have the time it takes to write it to disk.  

Current examples already push these limits into the realm of big data. High-resolution climate modeling (e.g., \citet{baker2014}) can generate 1 terabyte (TB) of data per compute day and run for 50 days on the NCAR-Wyoming supercomputer Yellowstone using 23,404 cores \citep{Small2014}. Investigations of the large scale spatial structure of the Universe (e.g., \citet{Heitmann2014}) produced 60 TB of data from nearly 1000 $N$-body simulations, taking a week or more per simulation run on Los Alamos National Laboratory's Coyote supercomputer. 

A standard strategy to reduce I/O requirements is to define a sampling rate for different variables and simulation components and save the simulation state only at evenly spaced time intervals according to that sampling rate \citep{baker2014}.
Here we describe an alternative method for reducing I/O requirements: instead of saving every $k$th time step of a simulation, we use an online and computationally lightweight statistical model to identify a reduced set of time steps to save, while the simulation is running. In doing so we take advantage of another strategy for reducing I/O requirements, called \insitu\/ analysis \citep{ahern2011}, which moves analyses (diagnostics) into the simulation itself, rather than first transferring the simulation results (prognostics) to storage for later analysis. We designed our method for \insitu\/ implementation.

To be useful in this high performance computing context, our method must take into account several constraints in terms of computational costs and storage/bandwidth. First, since we're embedding our analyses in the simulation, we must limit our method's computational burden to avoid slowing down the simulation's progress. Second, because we don't have sufficient storage or bandwidth to save and then post-process the entire simulation, our method must operate in an online fashion, making greedy decisions at each time step as the simulation progresses without knowledge of future calculations. Third, these same storage and bandwidth limitations mean that we can't retain all the earlier simulation calculations up to the current time step, and so our method must make decisions based on only a summary of the past behavior of the simulation plus a few recent time steps that can be retained in and accessed by simulation memory. 

With these constraints in mind, we seek computationally lightweight analyses that allow us to do the following:
\begin{enumerate}
\item identify important elements of the simulation (e.g., important time steps or variables or spatial regions),
\item significantly reduce the amount of data we need to move and store in order to preserve these elements, and
\item facilitate future exploration of the stored reduced data.
\end{enumerate}
We note that scientists might not know in advance what constitutes an ``important'' element of the simulation, and so in this paper we focus on identifying  time steps where the simulation's behavior is changing. We provide a computationally simple screening tool to identify candidate time steps to explore. Our approach uses buffers and efficient update schemes that make it viable for use in large-scale
simulation environments. 

In \sect{sec:LCROSS} we introduce a large but still tractable computer simulation that we use for development and as a case study. As we describe in \sect{sec:method} and demonstrate in \sect{sec:demo}, we use piecewise linear modeling of the simulation output as a function of time, coupled with hypothesis testing, to choose breakpoints in the piecewise model in an online fashion as each time step is computed by the simulation. Both the estimation of the linear fits and the hypothesis test can be inexpensively computed using a small number of sufficient statistics that are tracked by simple accumulation. The time steps at the selected breakpoints can then be saved, along with the sufficient statistics capturing the linear fits, using a fraction of the space required to store the entire simulation time series. The saved discrete time steps can later be analyzed on their own or used in conjunction with the sufficient statistics to reconstruct a piecewise linear approximation of the entire simulation. The sufficient statistics also provide a measure of the quality of this reconstruction by way of the residual sum of squares, which will depend on the variability in the simulation and the settings of our algorithm's tuning parameters as discussed in  \sect{sec:tuning}.

Again, our choices for modeling and decision-making are dictated by the need to incur as little computational expense as possible while still providing value to the scientists. We present a simple approach because, in this regime, we can only afford simple approaches. Our goal is not to create a new statistical methodology or to precisely emulate the simulation output. Rather we demonstrate an innovative, statistically motivated procedure that can be embedded in the simulation itself to facilitate scientific discovery. As we discuss in \sect{sec:discussion}, this simple method both achieves the goals stated above and provides an excellent starting point for many lines of future research.

\subsection*{Related work}
Our work touches a number of well-studied areas in the statistical literature:  piecewise linear modeling, change point detection and process control, and computer model emulation.

Our method uses a sequential piecewise linear regression model (see, e.g., Chapter 5 of \cite{HastieTibshiraniFriedman2009} for a brief overview).  As we describe in \sect{sec:method}, we repeatedly choose between a single regression line or a pair of regression lines to describe the behavior of a sequence of simulation time steps.  The two-line model has been called {\em two-phase regression} in much previous work.  \citet{hinkley1971inference} describes estimation and inference for this model under the assumption that the regression lines intersect at a particular point.  \citet{worsley1983testing} describes two-phase regression with hypothesis testing that is similar in spirit to ours.  In both of these papers, the authors assume that all of the data is available for consideration, whereas we are faced with the need to make greedy local decisions without knowing all of the future data or retaining all of the past data.  \citet{joo2009jump} consider a similar problem where the decision making doesn't rely on complete data.  Their model is more general than ours, and the overall method is correspondingly more complicated, which presents difficulties in our setting where the computational expense needs to be kept very low. \citet{koo1997spline} presents an example of determining good knot points in a general modeling setting; our piecewise linear models can be thought of as a special case of this. Again, our situation differs because we need to make decisions without knowing future data.

Our scheme looks for changes in the behavior of a system as data become available, a problem often called {\em change point detection}.  The change point literature is vast, so we will focus on the literature concerning change point detection in piecewise linear models.  \cite{BaiPerron1998} discuss a general framework for change point detection in linear models, while   \cite{WangZivot2000} present a Bayesian method in a similar setting.  Both assume access to the entire series of data, and neither is constrained by the need for fast computations.  \cite{KeoghChuHartPazzani2001} present a review and new methodology for online segmentation of time series using linear segments.  This work is closely related to what we present in that they consider a sliding buffer of data and make decisions about breakpoints within this buffer.  Their reviewed algorithms either assume that all of the data is available or that an arbitrarily large amount can be buffered.  Their new algorithm assumes a buffer that seems to be much larger (able to contain five or six line segments) than those we consider feasible for our applications.  

Change detection is also the goal of statistical process control (e.g., \citet{qiu2013introduction}). The recent work presented by \citet{qiu2014univariate} is similar to the method we present here.  A major difference is that their fitting procedure has access to more data than we do.  In general, the process and quality control literatures provide a rich number of tests that may ultimately be adapted to the setting that we are considering.  For the present, we will consider a very simple test that can be computed using lightweight sufficient statistics, and future work will consider tests from this literature that may be beneficial.

Much of the literature on computer models focuses on approximating the output of a complex simulation at untried input settings.  
The work presented here is closest to that described in \cite{lawrence2010} where the focus is approximation of complex output of individual simulator runs.  Unlike our online approach that analyzes a single run of a simulation, their work is conducted offline and can consider the output of many runs to determine the best approximation in a particular model class.

\section{An example computer simulation: LCROSS} \label{sec:LCROSS}
To provide a concrete case study, we introduce a computer simulation run by \citet{korycansky2009} in support of NASA's 2009 Lunar Crater Observation and Sensing Satellite mission (LCROSS). LCROSS  operated in tandem with NASA's Lunar Reconnaissance Orbiter (LRO) mission to the Moon. Briefly, the goal of the LCROSS mission was to identify and characterize subsurface water on the Moon. It  accomplished this by directing a spent rocket from LRO into the Moon to impact the Moon's surface and generate a plume of ejecta. LCROSS followed the impact with a shepherding spacecraft to measure the content of the plume. Prior to the mission's launch, \citet{korycansky2009} used several complex computer models to predict characteristics of the impact crater and ejecta, such as size and mass. 

\begin{figure}
\begin{center}
\begin{tabular}{ccc}
$t_5$ & $t_{500}$ \\
\includegraphics[trim = 2.7in 1.25in 2.7in 1.25in, clip, width=2.5in]{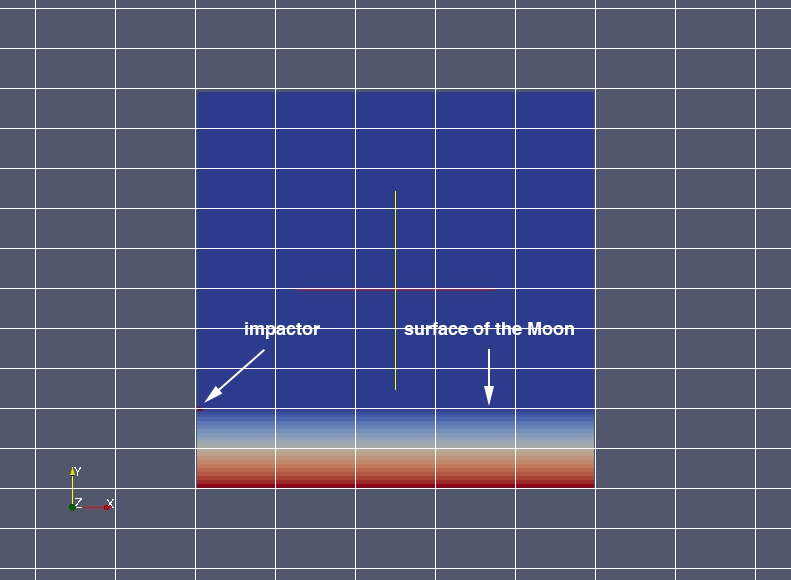} &
\includegraphics[trim = 2.7in 1.25in 2.7in 1.25in, clip, width=2.5in]{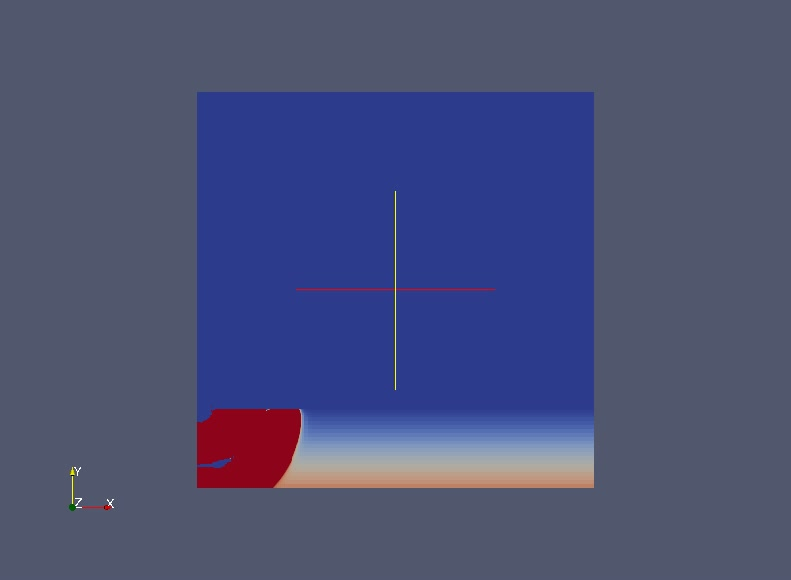} \\
$t_{1000}$ & $t_{2000}$ \\
\includegraphics[trim = 2.7in 1.25in 2.7in 1.25in, clip, width=2.5in]{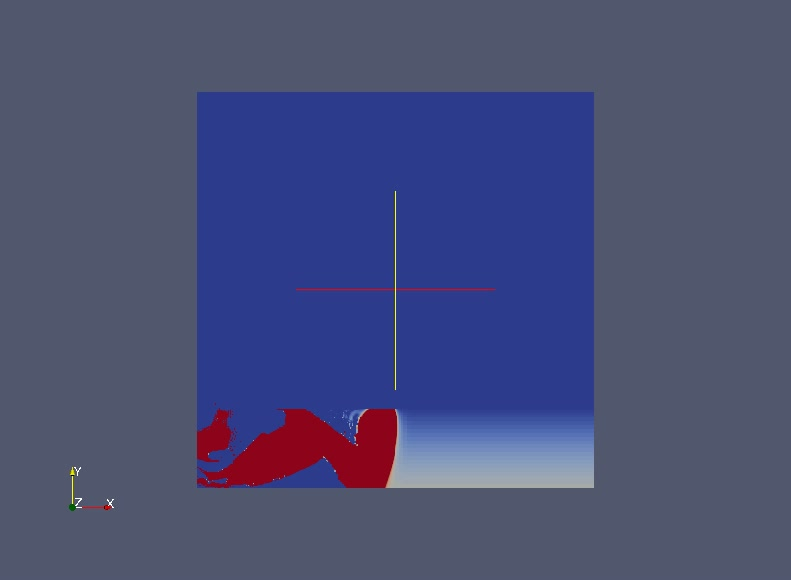} &
\includegraphics[trim = 2.7in 1.25in 2.7in 1.25in, clip, width=2.5in]{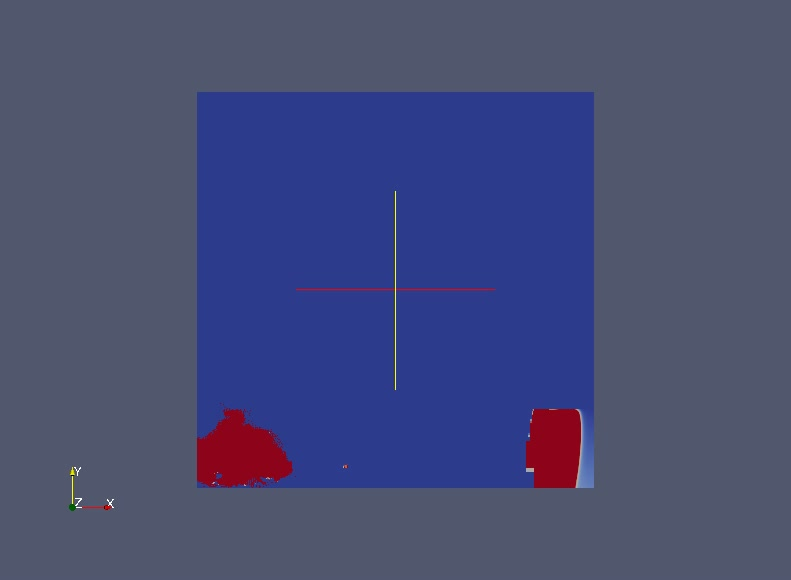}
\end{tabular}
\caption{Pressure variable from the 2D RAGE simulation of LCROSS at 4 time steps $t_i$, rendered with the ParaView visualization tool for large data sets \citep{ahrens2005}. The crosshairs in each panel are local spatial axes overlaid by ParaView. In the $t_5$ panel, we have drawn blocks to indicate contiguous regions within which we might perform individual analyses; see \fig{fig:prsBlocks} below. The surface of the Moon can be seen in the bottom two rows of blocks; the dark blue in the rows above that is atmosphere. The impact occurs at the extreme left edge of this panel, with the impactor shown as a small red spot in the top-left corner of its block. The remaining panels show how pressure evolves through the surface of the Moon after impact, with red indicating higher pressure. The entire simulation of 2000 time steps captures a time period of $< 0.1$ seconds after impact.}
\label{fig:LCROSS}
\end{center}
\end{figure}

We focus here on their use of a radiation-hydrodynamics code called RAGE, for Radiation Adaptive Grid Eulerian \citep{gittings2008}. RAGE is a massively parallel Eulerian code used to solve 1D, 2D, or 3D hydrodynamics problems. It uses an adaptive mesh refinement (AMR) algorithm to focus computations at each time step on the spatial regions with the most variation. \citet{korycansky2009} used RAGE to generate 2D predictions about the initial thermal plume that would occur within the first tenth of a second of the impact of the spent rocket into the Moon. \fig{fig:LCROSS} shows four of the 2000 time steps calculated for the pressure variable of the simulation. 

\begin{figure}
\begin{center}
\includegraphics[width=6in]{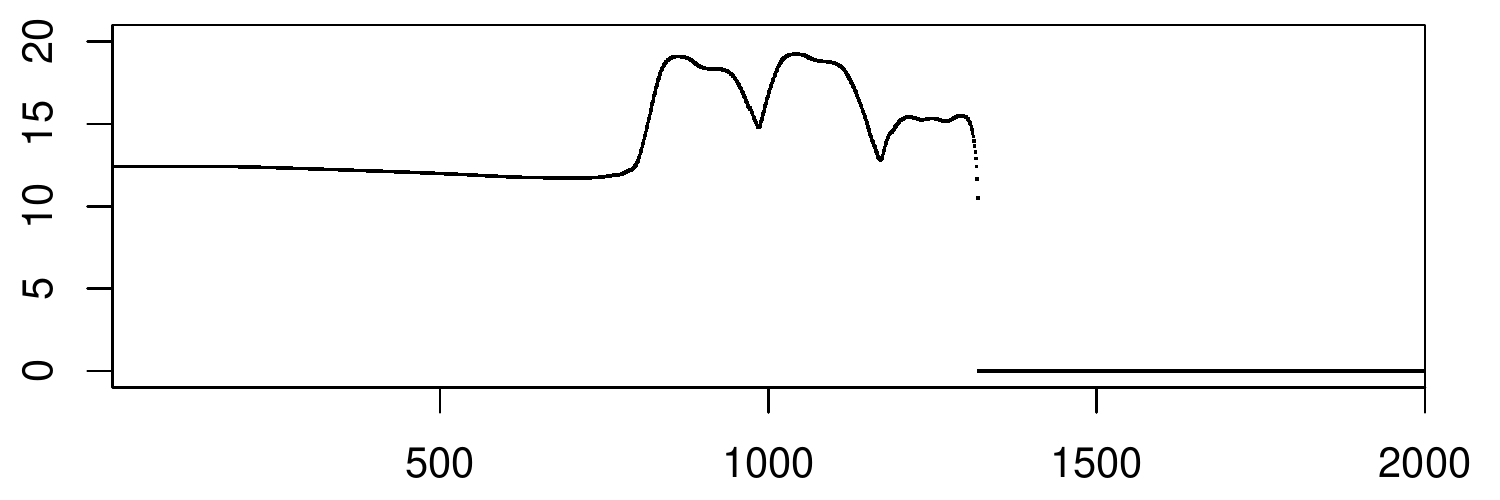}
\caption{Log of the LCROSS pressure variable for one grid point over 2000 simulation time steps. Note that this sequence of time steps is both highly correlated (lag one autocorrelation $> 0.99$) and nonstationary. These characteristics are typical of the deterministic computer simulations of interest here.}
\label{fig:prsPixel}
\end{center}
\end{figure}

\citet{korycansky2009} used 128 EV68 1.25 GHz processors in a supercomputing cluster at Los Alamos National Laboratory. While not an actual exascale simulation, this example provides a useful case study for developing and demonstrating our methods. We will focus most of our discussion on a 1D sequence extracted from the simulation, shown in \fig{fig:prsPixel} and representing the evolution of the pressure variable over 2000 simulated time steps at a single grid point. 

\fig{fig:prsPixel} demonstrates a number of key characteristics that make it both typical of the deterministic simulations of interest and challenging in terms of applying standard statistical approaches. The data in \fig{fig:prsPixel} are not independent or identically distributed but rather are highly correlated with a lag one autocorrelation greater than 0.99. Nor are they stationary:  the early part of the simulation is very flat (the first 500 time steps have a marginal variance of 0.02), the middle region shows greater variation (time steps 800 through 1300 have a marginal variance of 3.48), and the end of the simulation is completely flat. In our expected use cases, we don't have the luxury of knowing in advance how and when the simulation's behavior will change over time.  
These characteristics influence the decisions we make in developing our method, described in the next section. 

\section{Identifying important time steps} \label{sec:method}

As discussed in \sect{sec:intro}, we want to identify a reduced set of ``important'' time steps at which to save output from a time-dependent simulation.  In this context, important time steps are those at which the behavior of the simulation changes.  We will make this idea more precise as we describe the model.  

\begin{figure}
\begin{center}
\includegraphics{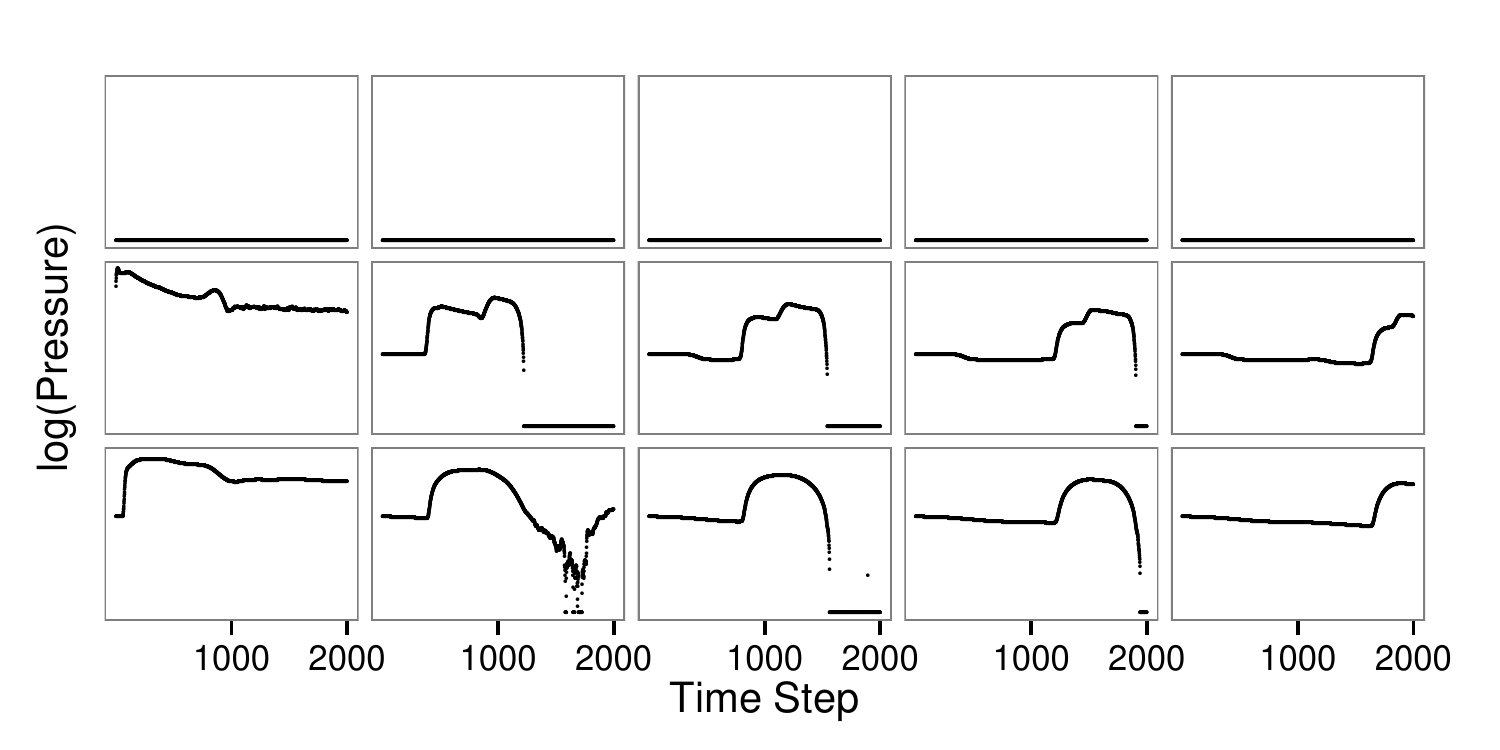}
\caption{Each panel shows the evolution over 2000 simulation time steps of the mean log pressure, where the mean is computed over a $50 \times 50$ block of grid points. These three rows of panels correspond to the bottom three rows of blocks shown in the $t_5$ panel of \fig{fig:LCROSS}. The bottom two rows are blocks completely within the surface of the Moon, while the top row shows blocks containing only the atmosphere immediately above the Moon, where pressure never changes due to impact in this simulation.}
\label{fig:prsBlocks}
\end{center}
\end{figure}

To characterize the behavior of the simulation, our approach considers a scalar response or a set of scalar responses $y_i$ that can be extracted or computed at each time step $t_i$. In the 2D LCROSS simulation we could consider the value of a single variable at one or more grid points in a regular grid overlaid on the AMR grid as seen in \fig{fig:prsPixel}; the mean of a variable across different spatial regions of interest in the simulation as shown in \fig{fig:prsBlocks}; or a derived quantity such as the correlation between two variables at a set of spatial locations. To maintain simplicity, we treat each response as if it were independent of the other responses, although this may not actually be true.

\subsection{Modeling the simulation} \label{model}
In a typical simulation setting, a scalar response $y_i$ will be an unknown deterministic function of time $t_i$:
\begin{equation}
\label{eq:F}
y_i = \mathcal{F}(t_i), ~ i=1,\ldots,T,
\end{equation}
where $T$ is the total number of time steps in the simulation.  Our goal is to approximate this function and identify important changes. We will approximate the function $\mathcal{F}$ with some other function $f$:
\begin{equation} 
\label{eq:fapprox}
y_i = f(t_i) + \epsilon_i, ~ i=1,\ldots,T,
\end{equation}
where $\epsilon_i$ captures the discrepancy between our approximation and the actual simulation.  As discussed in \sect{sec:LCROSS}, we do not expect these $\epsilon_i$ to be i.i.d.\ or even random.  This characteristic will influence the form of our hypothesis test, introduced in \sect{algorithm}.

As explained in \sect{sec:intro}, our choice of an approximating model in \eqref{eq:fapprox} must be computationally efficient to estimate so that it can be embedded in the simulation. It also can't rely on having access to either future time steps or past time steps in order to make decisions; it can only access a few recently calculated time steps of the simulation. In addition, we prefer a model in which changes are easy to define and detect.  


We therefore choose $f$ from the class of piecewise simple linear fits (e.g., \citet{HastieTibshiraniFriedman2009}, Chapter 5). 
Let $P_0, P_1, \ldots, P_m$ be a set of breakpoints of the sequence $1, \ldots, T$, with $P_0 = 0$ and $P_m=T$.  The function $f$ can be written as a sum over the partitions defined by the breakpoints:
\begin{equation} 
\label{eq:piecewise}
f(t_i) = \sum_{j=1}^{m} \left( \beta_{j,0} + \beta_{j,1} t_i \right) I\{P_{j-1} < i \le P_j \},
\end{equation}
where $\beta_{j,0}$ and $\beta_{j,1}$ are the intercept and slope in each partition $j$, and $I\{ \cdot \}$ is an indicator function.  This model says that there are $m$ non-overlapping partitions in $1, \ldots, T$, and that the function is linear within each partition.  To fit the model, we will need to estimate the number of partitions $m$, the breakpoints $P_j$, and the coefficients $\beta_{j,0}$ and $\beta_{j,1}$ of the linear fits. 

These linear fits are not only fast to compute, but they also can be efficiently updated via sufficient statistics as we describe in \sect{details}.  Furthermore, they allow us to define a change in behavior as a change in the piecewise linear fit, determined by a hypothesis test defined in \eqref{eq:hypotheses} below.
By only saving sufficient statistics for the simulation when triggered by the breakpoints between linear fits, we achieve substantial memory savings compared with storing the full output of the simulation. In addition, by saving these sufficient statistics we can later reconstruct a linear approximation of the simulation, and we can compute the residual sum of squares (RSS) of this reconstruction as a measure of its fidelity to the simulation. As we will demonstrate in \sect{sec:demo}, we gain improved fidelity to the simulation compared to the standard practice of saving evenly spaced time steps. 

\subsection{Estimating the piecewise linear model} \label{algorithm}
\begin{figure}
\begin{center}
(a) $t_{26}$: fail to reject single line fit \\
\includegraphics[width=4in]{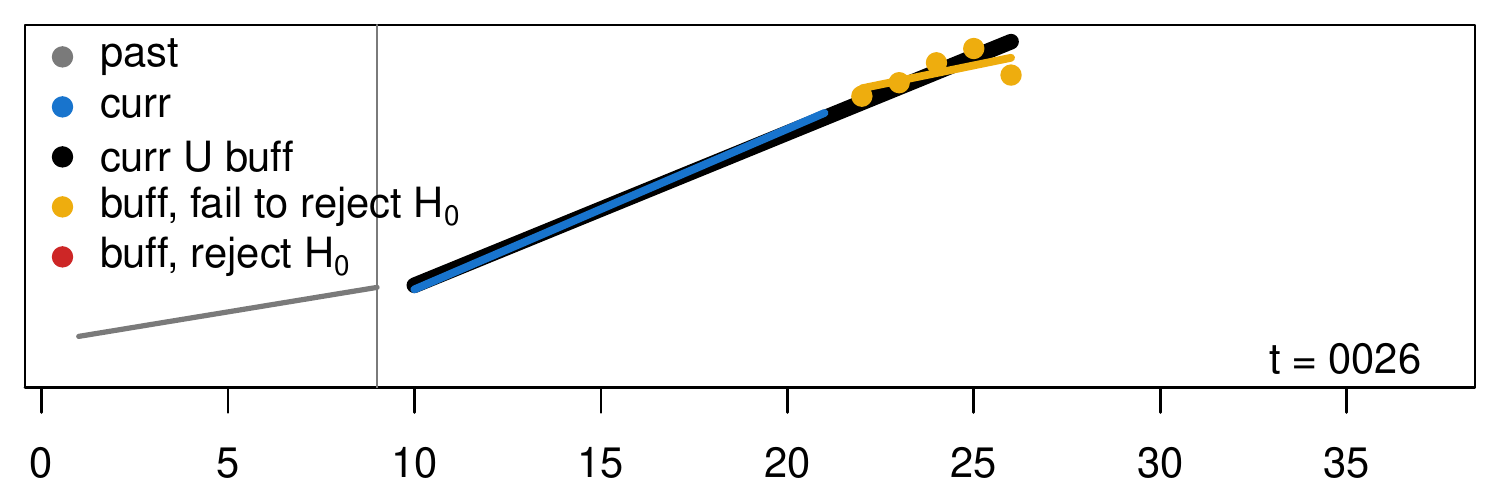} \\
(b) $t_{27}$: \curr{} $\leftarrow$ \curr{} $\cup$ {\sf old}, 
\buff{} $\leftarrow$ (\buff{}$\setminus${\sf old}) $\cup$ {\sf new}, reject single line fit \\
\includegraphics[width=4in]{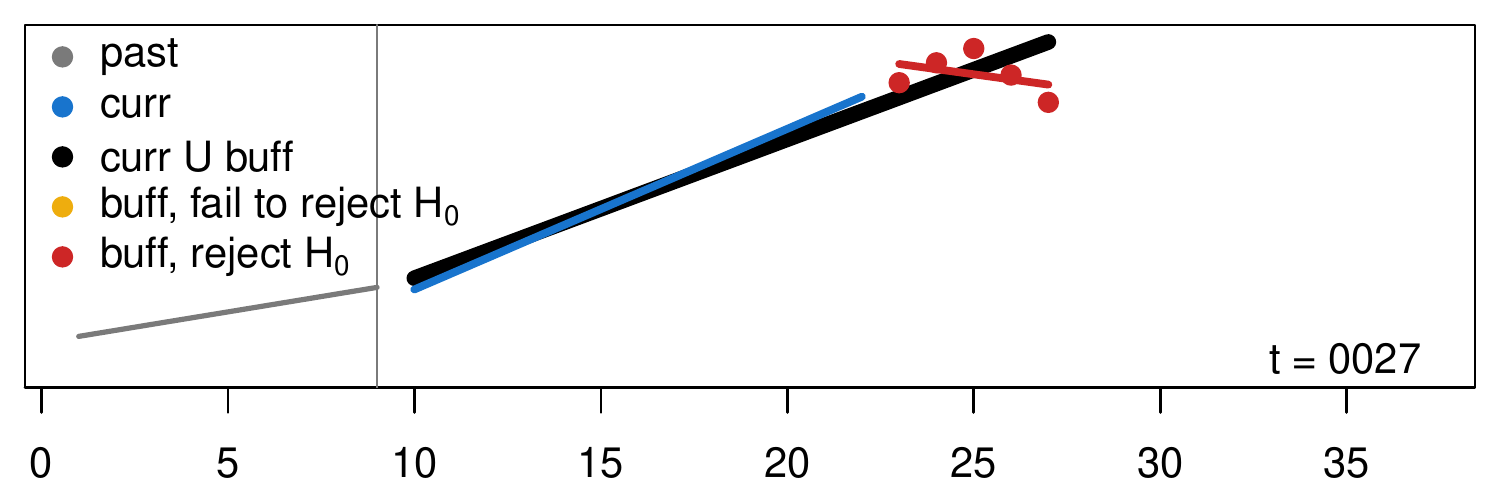}\\
(c) $t_{27}$: \past{} $\leftarrow$ \past{}  $\cup$ \curr{}, \curr{} $\leftarrow$ \buff{}\\
\includegraphics[width=4in]{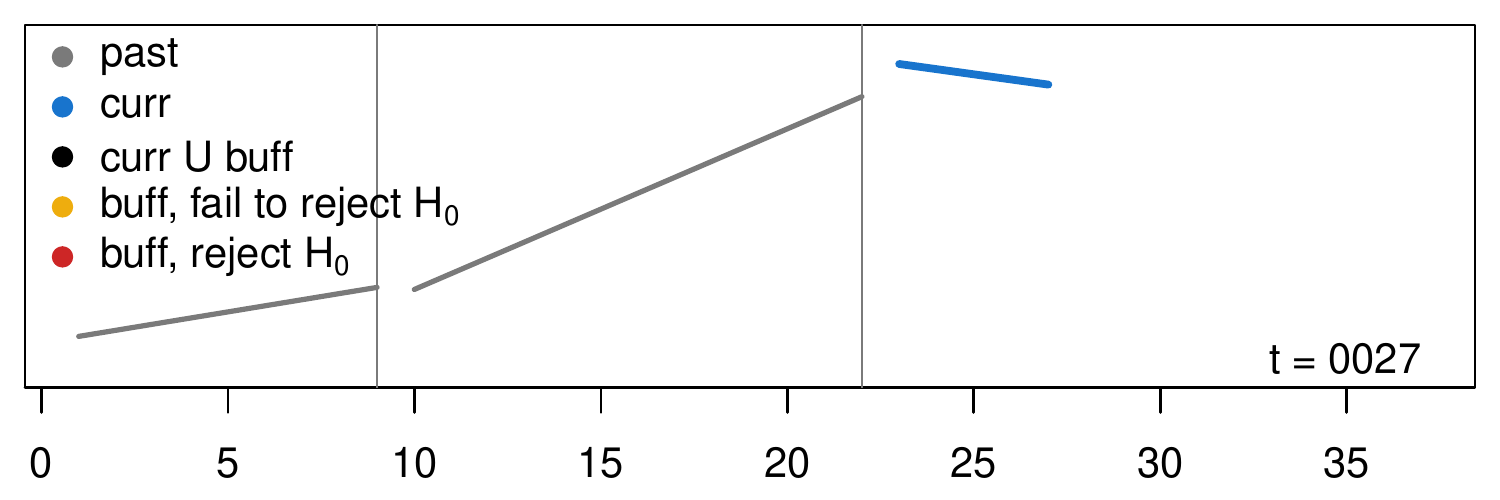}\\
(d) $t_{27 + B} = t_{32}$: \buff{} $\leftarrow$ next $B$ time steps\\
\includegraphics[width=4in]{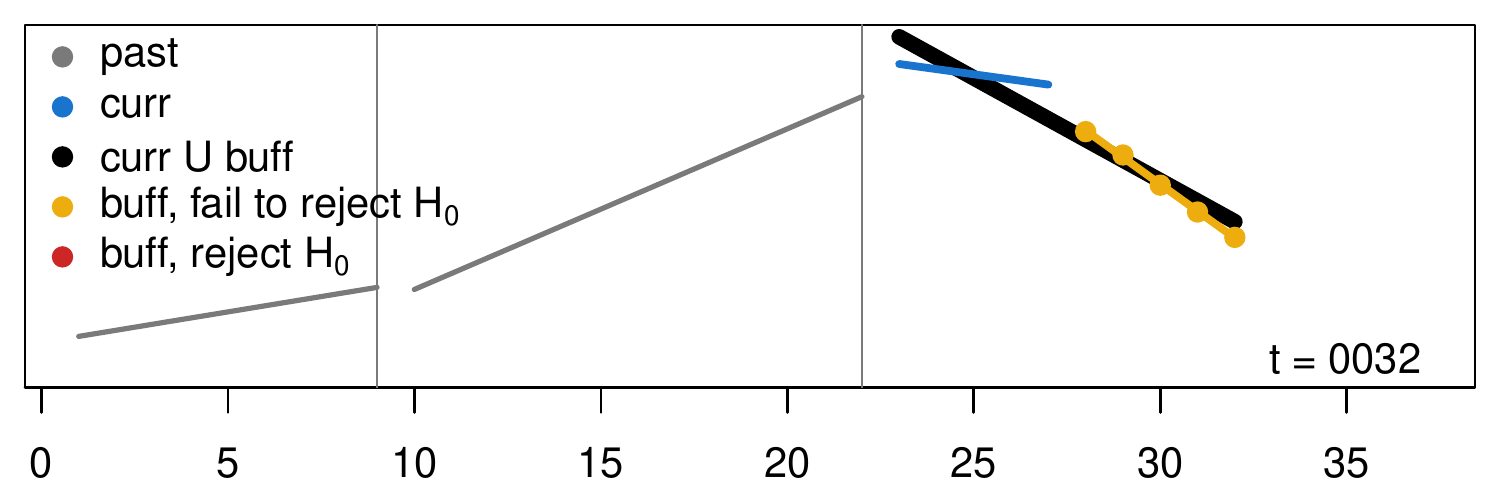}
\caption{Demonstration of the method with $B=5$ on simulated piecewise linear data with noise. Note that the \past{} (gray) and \curr{} (blue) regions are represented by lines only; the time steps in those regions are not saved due to memory constraints. Only the $B$ time steps in \buff{} are retained in memory. In (a) we compare the single line fit (black) to the fit of two lines (blue + yellow) and fail to reject $H_0$. In (b) we have moved the oldest time step in \buff{} to \curr{}, updated the line for \curr{}, and added a newly computed time step to \buff{}. This triggers a rejection of $H_0$. In (c) we mark the breakpoint, move \curr{} to \past{} and \buff{} to \curr{}, saving only their sufficient statistics. In (d) we assign $B$ newly computed time steps to \buff{} and continue the procedure.}
\label{fig:demo}
\end{center}
\end{figure}

Here we introduce our greedy online approach for estimating the model in \eqref{eq:piecewise}. We assume that we have the computational resources to maintain $B$ time steps in a buffer. These $B$ time steps, but no others, are accessible by our embedded \insitu\/ method. 
At each time step we divide the simulation into three temporal regions as shown in \fig{fig:demo}:
\begin{enumerate}
\item \buff{}: $B$ time steps newly computed by the simulation and stored in the buffer. \fig{fig:demo} shows these in yellow or red, depending on whether those $B$ time steps lead us to reject our hypothesis and choose a new breakpoint. These are the only time steps stored in simulation memory and therefore available to our method.
\item \curr{}: Time steps currently characterized by a linear fit and therefore no longer stored in simulation memory. These are shown as blue lines in \fig{fig:demo}.
\item \past{}: Older time steps no longer under consideration (and no longer stored in memory) because we've already captured them via their breakpoints and sufficient statistics. These are shown as gray lines in \fig{fig:demo}.
\end{enumerate}
For notational convenience, we additionally define {\sf old} to be the oldest time step stored in \buff{}, and we let {\sf new} denote the next new time step to be computed by the simulation.

At each simulation time step, our method compares the time steps in \curr{} and \buff{}. The newest time step in \curr{} is a candidate breakpoint. We consider the following hypotheses:
\begin{equation}
\begin{aligned}
H_0 &: \text{A single line provides the best fit for \curr{}} \cup \text{\buff{}}. \\
H_1 &: \text{A two-line fit (one line for \curr{}, one for \buff{}) provides the best fit.}
\end{aligned}
\label{eq:hypotheses}
\end{equation}
In this greedy method of estimating the model, rejecting $H_0$ immediately places a breakpoint between \curr{} and \buff{} and defines \curr{} as one of the partitions in \eqref{eq:piecewise}. 

\fig{fig:demo} walks through the method in a synthetic example of piecewise linear data with Gaussian noise (which of course is not what we expect in practice but which provides a useful illustration). In \fig{fig:demo}~(a) we have already chosen a breakpoint at $t_{9}$, indicated by a vertical line, and we are now processing later time steps in the simulation. At $t_{26}$ in (a) we fail to reject $H_0$ after comparing \curr{} and \buff{}. That is, we fail to reject the single-line fit (black line) in favor of a two-line fit (blue + yellow lines). We therefore progress in the simulation by moving {\sf old} into \curr{} and adding {\sf new} to \buff{}. As seen in \fig{fig:demo}~(b), {\sf old} is no longer retained as a time step in memory and is only used to update the blue line for \curr{}. For $t_{27}$ in \fig{fig:demo}~(b), this update leads us to reject $H_0$. In \fig{fig:demo}~(c) we set a breakpoint at the last time step in \curr{}, move \curr{} to \past{}, and save its sufficient statistics (and possibly some component of the prognostic simulation output if we have sufficient storage and bandwidth available). We also move \buff{} to a new \curr{}, saving only the sufficient statistics (not the individual time steps). In \fig{fig:demo}~(d) we compute the next $B$ time steps of the simulation and add them to \buff{}, then continue the algorithm.

In this example, note that the second breakpoint is placed at $t_{22}$ rather than at the true underlying breakpoint at $t_{25}$. This behavior is due to the fact that our greedy online method treats the time steps in \buff{} as a unit rather than as a collection of individual time steps. As we will see in \sect{sec:tuning}, our need to make greedy decisions like this can have a strong local impact on the fit --- that is, it can lead to a high RSS for a particular partition.  If we had sufficient computational resources, we could consider investigating each time step in \buff{} to determine whether it should be a breakpoint, but we don't have that luxury here. However, our method also provides a means of measuring the quality of the fit of each partition via the RSS, so that a scientist could return to regions of the simulation with high RSS to see what caused the poor fit.

\subsection{Computational details} \label{details}
Table~\ref{tab:algorithm} outlines our algorithm. At each simulation time step we test the hypotheses in \eqref{eq:hypotheses} to compare the single- and two-line fits. To do this we compute the RSS for three regression lines --- \curr{}, \buff{}, and \curr{}~$\cup$~\buff{} --- which we can do efficiently using the following sufficient statistics for each line:
\begin{equation}
\theta  = \sum t_i, \;
\Theta  = \sum t_i^2, \;
\psi  = \sum y_i, \;
\Psi  = \sum y_i^2, \; \text{and}\;
\tau  = \sum t_i y_i,
\label{eq:suffstats}
\end{equation}
as well as $T_\bullet$, the number of time steps included in the regression line. All the sums in \eqref{eq:suffstats} are taken over the $T_\bullet$ time steps. Then we can compute the RSS for a particular regression line as follows:
\begin{equation}
RSS = \Psi - \frac{1}{T_\bullet} \psi^2 - \frac{(\tau - \theta \psi / T_\bullet)^2}{\Theta - \theta^2 / T_\bullet}.
\label{eq:RSS}
\end{equation}

Typically one would use an $F$-statistic to make the comparison. However, as discussed in \sect{sec:LCROSS}, the output of the deterministic computer simulations of interest here violates most of the usual assumptions we make when using a statistical test. We will therefore use a modified version of the $F$-statistic, which we define in \sect{sec:modF} below. 

\begin{table}
\begin{tabular}{p{.01\textwidth}p{.01\textwidth}p{.8\textwidth}}
\hline
{\bf Initialization:}\\
\hline
& \multicolumn{2}{l}{--  Set the buffer size $B$, choose values for $\alpha$ and $\delta^2$.} \\
& \multicolumn{2}{l}{--  Set the number of parameters in the single- and two-line models to $p_1 = 2$, $p_2 = 4$.} \\
& \multicolumn{2}{l}{--  \curr{} $\leftarrow$ the first $B$ time steps in the simulation.} \\
& \multicolumn{2}{l}{--  \buff{} $\leftarrow$ the next $B$ time steps in the simulation.}  \\
& \multicolumn{2}{l}{--  Compute the sufficient statistics for \curr{}, \buff{}, and \curr{} $\cup$ \buff{} via \eqref{eq:suffstats}.} \\
\hline
\multicolumn{3}{l}{\bf For each time step while the simulation is running:}\\
\hline
& \multicolumn{2}{l}{--  Compute $RSS_{\sfsub{curr}}$, $RSS_{\sfsub{buff}}$, and $RSS_1 \equiv RSS_{\text{\curr{}} \cup \text{\buff{}}}$ via \eqref{eq:RSS}.}\\
& \multicolumn{2}{l}{--  $RSS_2 \leftarrow RSS_{\sfsub{curr}} + RSS_{\sfsub{buff}}$.}\\
& \multicolumn{2}{l}{--  Compute a modified $F$-statistic via \eqref{eq:Ftestmod}, compare to $\alpha$.}\\
& \multicolumn{2}{l}{--  If we reject $H_0$:}\\
&& -- Save the sufficient statistics for \curr{} and (optionally) the simulation output around the breakpoint. \\
&& --  \past{} $\leftarrow$ \past{} $\cup$ \curr{}.\\
&& --  \curr{} $\leftarrow$ \buff{}.\\
&& --  \buff{} $\leftarrow$ the next $B$ time steps in the simulation.\\
&& --  Compute the sufficient statistics for \curr{}, \buff{}, and \curr{} $\cup$ \buff{} via \eqref{eq:suffstats}. \\
& \multicolumn{2}{l}{--  Else:}\\
&& --  \curr{} $\leftarrow$ \curr{} $\cup$ {\sf old}.\\
&& --  \buff{} $\leftarrow (\text{\buff}\setminus\text{{\sf old}}) \cup \text{{\sf new}}$.\\
&& --  Compute the sufficient statistics for \buff{} via \eqref{eq:suffstats}. \\
&& --  Update the sufficient statistics for \curr{} and \curr{} $\cup$ \buff{} via \eqref{eq:updates}. \\
\hline
\end{tabular}
\caption{A greedy online algorithm for estimating the piecewise linear model in \eqref{eq:piecewise}.}
\label{tab:algorithm}
\end{table}

At each new time step, the sufficient statistics for \buff{} can be recomputed inexpensively via \eqref{eq:suffstats} using the $B$ time steps currently stored in the buffer. We can also inexpensively update the sufficient statistics for \curr{} and \curr{}~$\cup$~\buff{} without retaining the original time steps in memory. Let $S$ be any of $\theta$, $\Theta$, $\psi$, $\Psi$, or $\tau$, i.e., any of the summation sufficient statistics.  Let $X_{\sf{old}}$ denote the  summand based on the oldest time step in \buff{} and let $X_{\sf{new}}$ denote a summand based on a time step newly computed by the simulation. The sufficient statistics for \curr{} and \curr{}~$\cup$~\buff{} are then updated as follows:
\begin{equation}
\begin{aligned}
S_{\text{\curr}}& \leftarrow S_{\text{\curr}} + X _{\sf{old}}\\
S_{\text{\curr} \cup \text{\buff}} & \leftarrow S_{\text{\curr} \cup \text{\buff}} + X_{\sf{new}}.
\end{aligned}
\label{eq:updates}
\end{equation}
The updates for $T_\bullet$ for each regression line are simple as well:  $T_\text{\curr}$ and $T_{\text{\curr} \cup \text{\buff}}$ increase by 1 while $T_\text{\buff}$ remains the same ($T_\text{\buff} = B$ always). 
Thus these updates and the subsequent computation of the RSS in \eqref{eq:RSS} have constant computational complexity at each simulation time step. We do not need to retain the simulation time steps themselves for the updates, just these sufficient statistics. 

Note that we don't need to estimate the regression coefficients in \eqref{eq:piecewise} in order to carry out the hypothesis test; we only need the RSS calculations in \eqref{eq:RSS}. However, for post-processing we do use the regression coefficients to reconstruct the simulation with our linear fits. We can compute the coefficients using the saved sufficient statistics for each partition $j$: 
 \begin{equation}
\begin{aligned}
\hat{\beta}_{j, 0} & = \frac{1}{T_\bullet}(\psi - \hat{\beta}_{j,1} \theta) \\
\hat{\beta}_{j, 1} & = \frac{\tau - \theta \psi / T_\bullet}{\Theta - \theta^2 / T_\bullet}.
\end{aligned}
\label{eq:betas}
\end{equation}

One concern with very long simulations, or those with long partitions, is that \eqref{eq:RSS} and \eqref{eq:betas} may produce numerical errors due to the size of the sums in \eqref{eq:suffstats}.  In this case, the algorithms described in \cite{Gentleman1974} and \cite{Miller1992} provide a numerically stable alternative for building and updating linear models. They use planar rotations for the updates and have the same memory requirements as our sufficient statistics algorithm in Table~\ref{tab:algorithm}.  Since the updates in these algorithms have the same complexity at each time step regardless of how many time steps have come before, they are well suited for an \insitu\/ implementation. The R package {\sf biglm} of \cite{biglm} provides an example of their implementation.

%
\subsection{A modified $F$-statistic} \label{sec:modF}
Let $RSS_1 \equiv RSS_{\text{\curr{}} \cup \text{\buff{}}}$ denote the residual sum of squares for the single-line fit, $RSS_2 = RSS_{\sfsub{curr}} + RSS_{\sfsub{buff}}$ the residual sum of squares for the two-line fit. The standard definition of the $F$-statistic is as follows:
\begin{equation}
\label{eq:Ftest}
F = 
\frac{\left(
	\frac{RSS_1 - RSS_2}{p_2 - p_1}
	\right)}
	{\left(
	\frac{RSS_2}{T_{\text{\curr{}} \cup \text{\buff{}}} - p_2}
	\right)},
\end{equation}
where $p_1$ and $p_2$ denote the number of parameters in the single- and two-line fits, and $T_{\text{\curr{}} \cup \text{\buff{}}}$ denotes the total number of time steps under consideration.
In a typical regression setting, $F$ has an $F$-distribution with $(p_2-p_1, T_{\text{\curr{}} \cup \text{\buff{}}} -p_2)$ degrees of freedom when $H_0$ holds. In that setting we would choose an $\alpha$ level that governs our desired probability of false rejection and use \eqref{eq:Ftest} to determine whether to reject $H_0$.

As we've discussed, the deterministic computer simulations we consider do not present the typical regression setting and in fact violate most assumptions made when using a statistical test. 
In practice, using the standard $F$-statistic at some $\alpha$ level with these computer simulations leads to unwanted rejections of $H_0$ when the RSS of \curr{} and \buff{} are both quite small, i.e., when the simulation is nearly linear.   
Also, the lack of a white noise error component and the high autocorrelation in these simulations mean that the values of $\alpha$ used in many standard settings (e.g., $\alpha = 0.05$) will lead to a very large overall number of partitions, which is not desirable with our limited bandwidth and storage.  We will demonstrate and discuss this behavior in \sect{sec:tuning}.

We therefore make a slight modification to the usual $F$-statistic to produce the behavior we desire, namely that we want our hypothesis test to ignore certain kinds of changes in the simulation. We add a new parameter, $\delta^2$, to the computed RSS for the single- and two-line fits, scaled by the total number of time steps under consideration, $T_{\text{\curr{}} \cup \text{\buff{}}}$:

\begin{equation}
\label{eq:Ftestmod}
F = 
\frac{\left(
	\frac{(RSS_1 + \delta^2 T_{\text{\curr{}} \cup \text{\buff{}}}) - (RSS_2 + \delta^2 T_{\text{\curr{}} \cup \text{\buff{}}})}{p_2 - p_1}
	\right)}
	{\left(
	\frac{RSS_2 + \delta^2 T_{\text{\curr{}} \cup \text{\buff{}}}}{T_{\text{\curr{}} \cup \text{\buff{}}} - p_2}
	\right)}
= 
\frac{\left(
	\frac{RSS_1 - RSS_2}{p_2 - p_1}
	\right)}
	{\left(
	\frac{RSS_2 + \delta^2 T_{\text{\curr{}} \cup \text{\buff{}}}}{T_{\text{\curr{}} \cup \text{\buff{}}} - p_2}
	\right)}.
\end{equation}
Adding this tuning parameter $\delta^2$ has the effect of inflating the estimate of the error variance in the two-line model so that the two-line fit needs to be much better than the single-line fit in order to reject the single-line model. As specified in Table~\ref{tab:algorithm}, we use \eqref{eq:Ftestmod} rather than \eqref{eq:Ftest} for our method. We emphasize that in our scenario we don't interpret $\alpha$ as a desired probability of false rejection but rather as one of our tuning parameters for assessing whether there is a change in behavior.

We will discuss the impact of $\alpha$ and $\delta^2$ on our algorithm in \sect{sec:tuning} and provide guidance for setting them. To give context for that discussion, we first demonstrate our method with the LCROSS data using some particular choices for $\alpha$ and $\delta^2$.

\section{Demonstration with the LCROSS simulation} \label{sec:demo}
\begin{figure}
\begin{center}
(a) Evenly spaced time steps: 25 partitions, $RSS_{\sf total} = 1140.15$ \\
\includegraphics[width=6in]{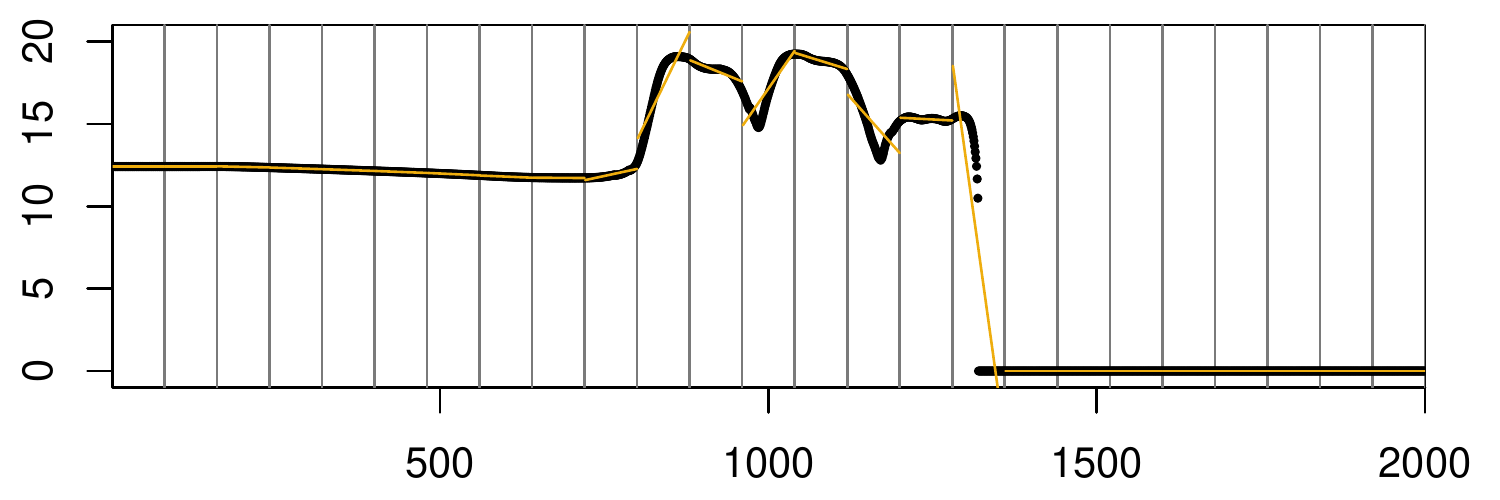}\\
(b) $\alpha = 0.001$, $\delta^2 = 0.001$: 25 partitions, $RSS_{\sf total} = 6.40$ \\
\includegraphics[width=6in]{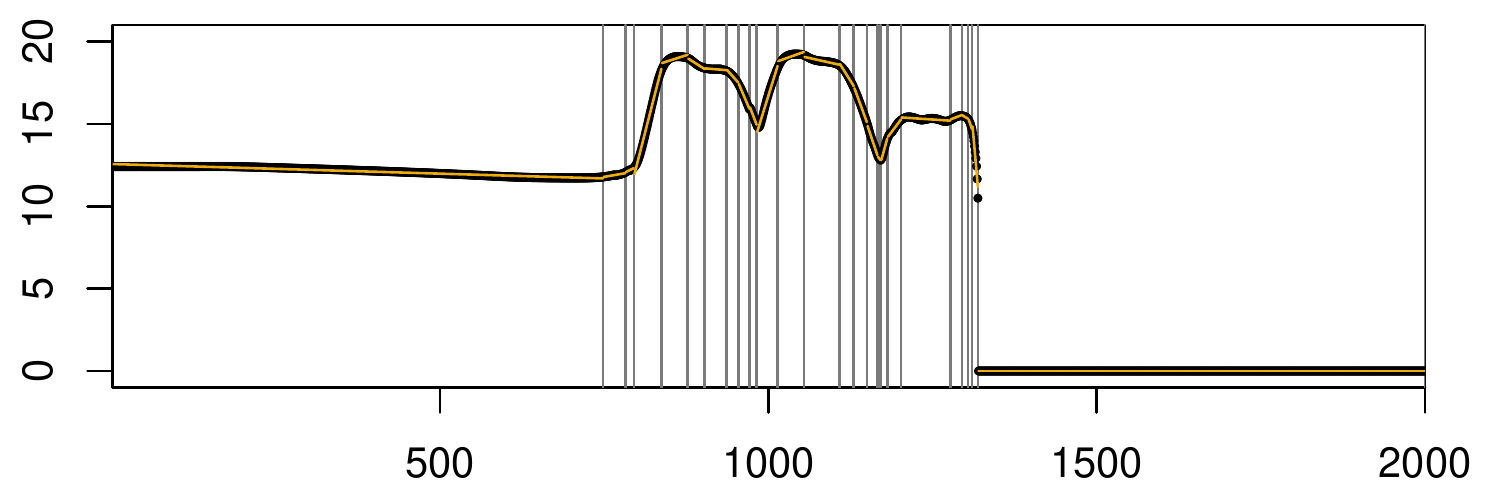}\\
(c) $\alpha = 10^{-7}$, $\delta^2 = 0.1$: 5 partitions, $RSS_{\sf total} = 1027.46$\\
\includegraphics[width=6in]{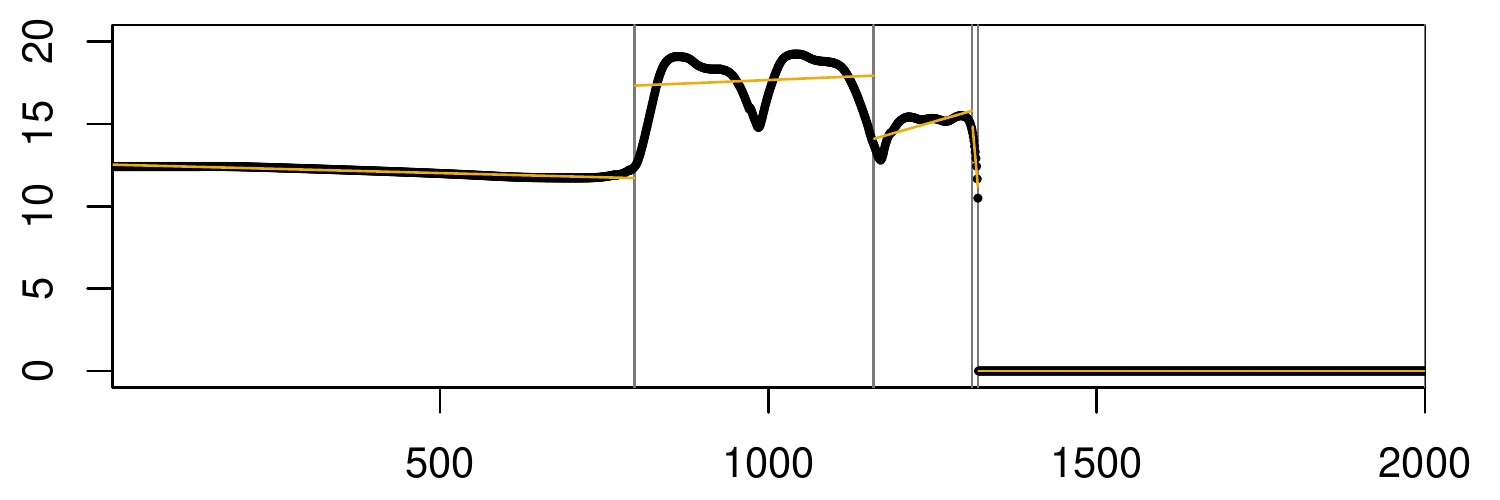}
\caption{Panel (a) shows 25 evenly spaced time steps, a common strategy for reducing the number of save points. Linear interpolation between these saved time steps (as shown in yellow) yields a fit with $RSS_{\sf total} = \sum_p RSS_p = 1140.15$. Panel (b) shows a setting of the tuning parameters $\alpha$ and $\delta^2$ that also produces 25 time steps while reducing $RSS_{\sf total}$ by 3 orders of magnitude. Panel (c) shows a setting that produces a similar $RSS_{\sf total}$ with only 5 partitions, a 5-fold savings.}
\label{fig:pixeldemo}
\end{center}
\end{figure}

We use the LCROSS simulation introduced in \sect{sec:LCROSS} to demonstrate the behavior of our method on a deterministic computer simulation. We first consider the log of the pressure variable for the single grid point shown in \fig{fig:prsPixel}. 
\fig{fig:pixeldemo} (a) shows the standard practice of saving evenly spaced time steps, here with 25 time steps saved (an arbitrary choice for this illustration). If we linearly interpolate between these saved time steps, $RSS_{\sf total} = \sum_p RSS_p = 1140.15$. As an alternative to these evenly spaced time steps, a scientist might want to use our method either to identify and save a similar number of time steps or to achieve a similar $RSS_{\sf total}$. Panel (b) shows a setting of the tuning parameters $\alpha$ and $\delta^2$ that produces 25 time steps and reduces $RSS_{\sf total}$ by 3 orders of magnitude. Panel (c) shows a setting that produces a similar $RSS_{\sf total}$ with only 5 partitions, a 5-fold savings. In both cases we use buffer size $B = 5$. 

In this example we have handpicked evocative settings of our tuning parameters $\alpha$ and $\delta^2$ to demonstrate the possibilities for significant improvements in fidelity to the simulation or significant reductions in storage and bandwidth requirements. With the relatively small LCROSS example we have the luxury of trying many different settings to explore the space --- indeed we will show 110 different settings in \sect{sec:tuning}. In practice we expect that the improvements over standard practice will be substantial even if the settings aren't ``optimized'' in some sense. We will illustrate the impact of the tuning parameters and discuss strategies for selecting them in \sect{sec:tuning}. In future work we will explore development of a mathematical framework for defining an optimal setting.

\begin{figure}
\begin{center}
\includegraphics{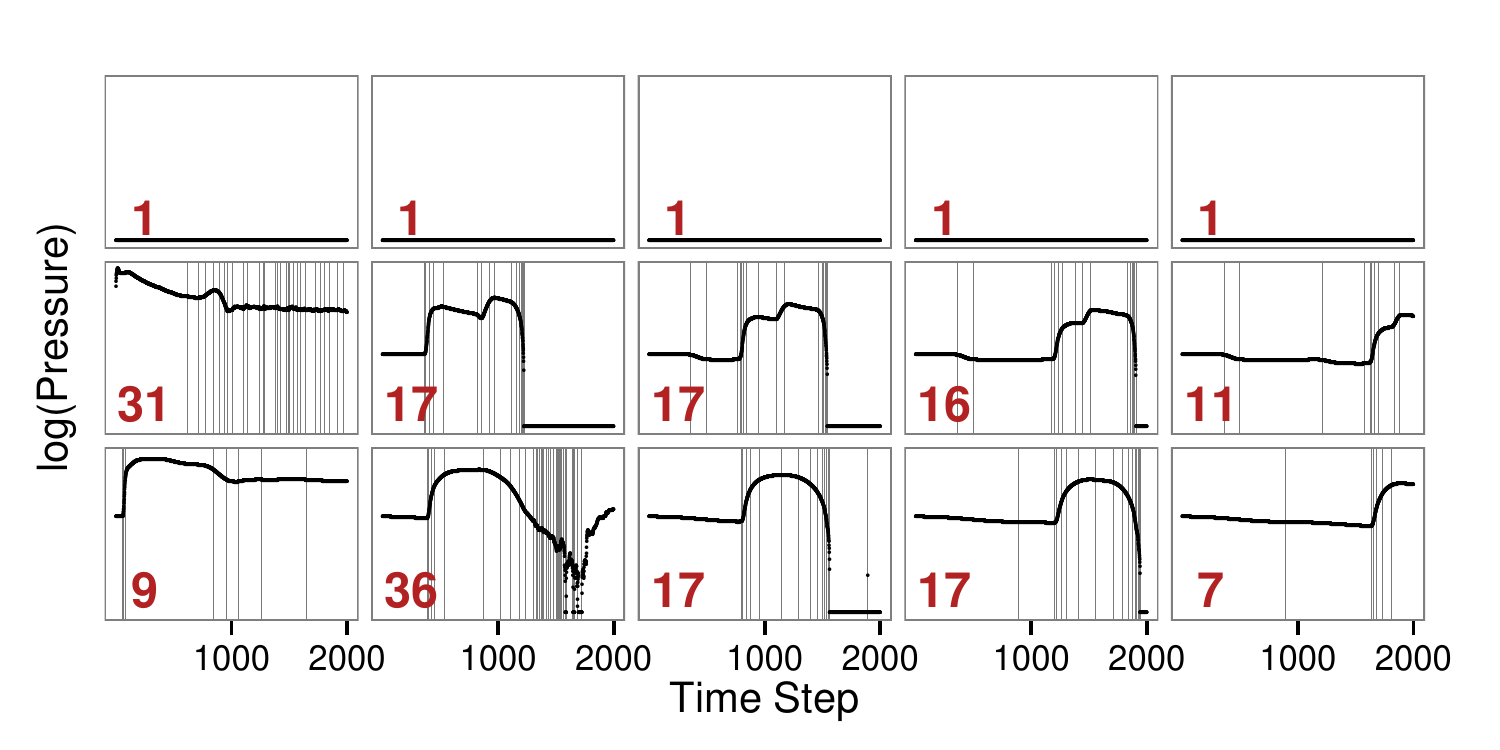}
\caption{Demonstration of the method on the block means shown in \fig{fig:prsBlocks} with $\alpha = 0.001$ and $\delta^2 = 0.001$. Vertical lines indicate identified breakpoints, and the number in each panel indicates the number of partitions selected. The top row of panels comprises grid points in the atmosphere above the Moon where no change in pressure is expected. The first panel in the middle row is where the spent rocket first impacts the Moon.}
\label{fig:prsBlocksdemo}
\end{center}
\end{figure}

Of course most simulations of interest will have a spatial component, either in 2D or 3D. 
A simple way to use our method to account for spatial characteristics is to reduce the spatial output to a set of 1D output as we did in \fig{fig:prsBlocks} by computing block means. 
\fig{fig:prsBlocksdemo} shows the result of applying our method to those block means with $\alpha = 0.001$, $\delta^2 = 0.001$, and $B = 5$. 
Again the vertical lines indicate the partitions selected, and the numbers indicate how many partitions were selected in each block. Recall that the top row of panels corresponds to blocks of grid points in the atmosphere above the Moon. The pressure variable in these atmosphere blocks is not affected by the impact, so a single partition comprising the entire simulation provides perfect fidelity in those blocks. Even in the most active blocks we select fewer than 40 partitions, representing a substantial savings while providing greatly improved fidelity. In future work we will explore other mechanisms for accounting for the spatial aspect of the simulations while still respecting our computational constraints.

\section{Setting the tuning parameters} \label{sec:tuning}
Recall that our algorithm has three tuning parameters: $B$, the buffer size; $\alpha$, which governs local comparisons of \curr{} and \buff{} in the hypothesis test \eqref{eq:hypotheses}; and $\delta^2$, which we introduced in \sect{sec:modF} to encourage less (or more judicious) rejection of $H_0$. Here we explore the behavior induced by different settings of these parameters and make some recommendations for how to choose them.

The buffer size $B$ can be selected based in part on the size of the data and the available memory, storage, and transfer resources. Preliminary timing studies may be helpful in arriving at a judicious value for $B$.
To estimate our piecewise linear fit in \eqref{eq:piecewise}, we need $B \ge 3$ so that we can estimate the two parameters of the line. Beyond that constraint, the choice for $B$ will be driven by characteristics of the available computing resources and the simulation itself. It will typically be chosen to reflect the desired tradeoff between computing costs and tolerance for uncertainty associated with the stored information. For the remainder of this section we will focus on $\alpha$ and $\delta^2$.

\begin{figure}
\begin{center}
\begin{tabular}{cc}
Number of Partitions & Total RSS (rounded)\\
\includegraphics[width=3in]{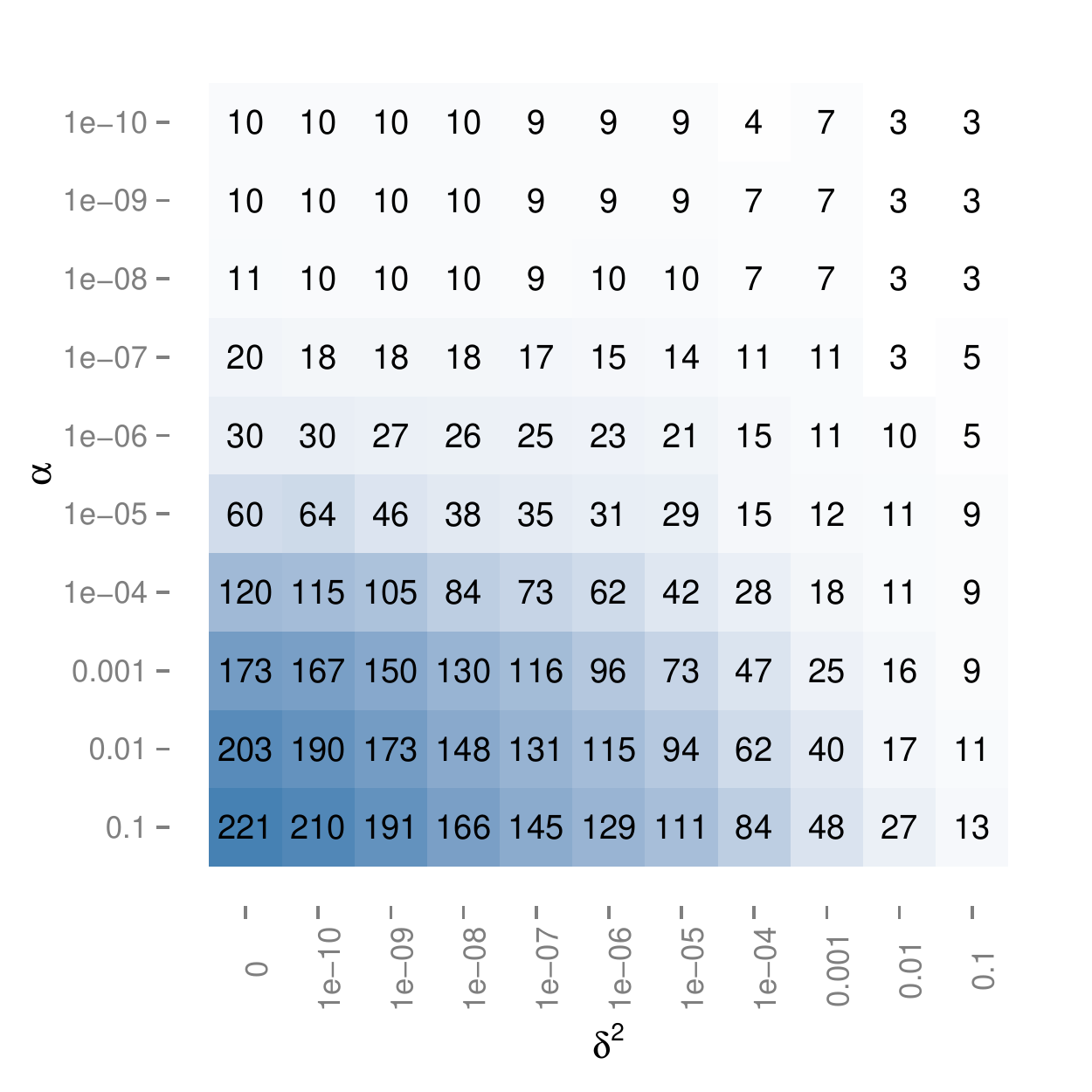} &
\includegraphics[width=3in]{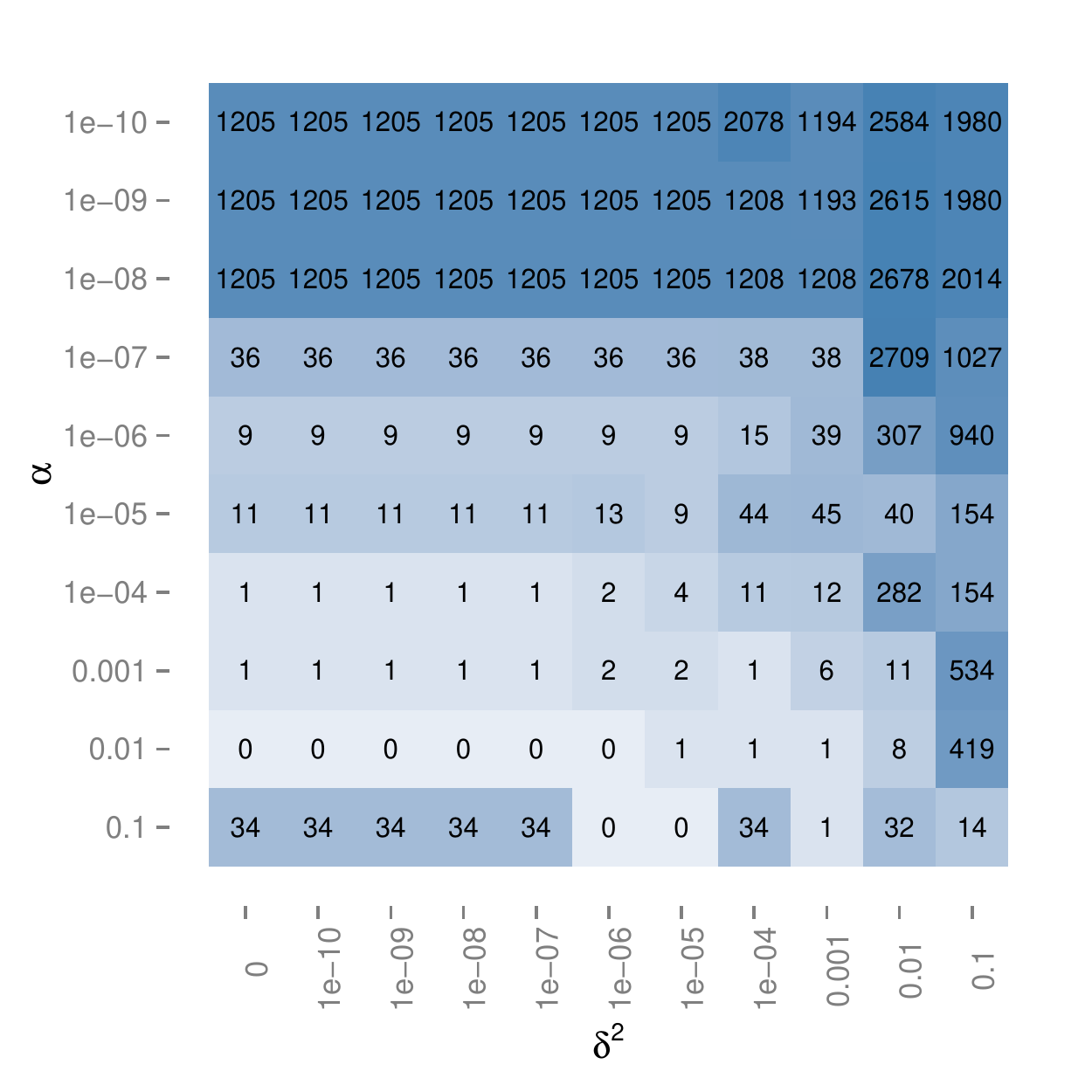}
\end{tabular}
\caption{Number of partitions selected (left) and the resulting $RSS_{\sf total}$ (right, rounded to the nearest integer) for a range of settings of $\alpha$ and $\delta^2$.}
\label{fig:heatmaps}
\end{center}
\end{figure}

The decisions made by our algorithm --- that is, the number and location of the partitions chosen --- will impact both the amount of storage and bandwidth required and the fidelity of the resulting reconstruction to the simulation. We can think about balancing the tradeoff between these aspects, i.e., achieving sufficient fidelity for a reasonable cost. Understanding the tradeoffs between uncertainty of a representation and computational cost is currently of paramount interest to the computing community \citep{whitaker2015}. Using the same grid point seen earlier in Figures~\ref{fig:prsPixel} and \ref{fig:pixeldemo}, we explore in \fig{fig:heatmaps} the number of partitions selected and the resulting total RSS for a range of settings of $\alpha$ and $\delta^2$. We include the case $\delta^2 = 0$ (first column of the two tables in \fig{fig:heatmaps}) to determine whether our introduction of $\delta^2$ in our method indeed provides benefit over simply changing $\alpha$.

\begin{figure}
\begin{center}
\includegraphics[width=6.5in]{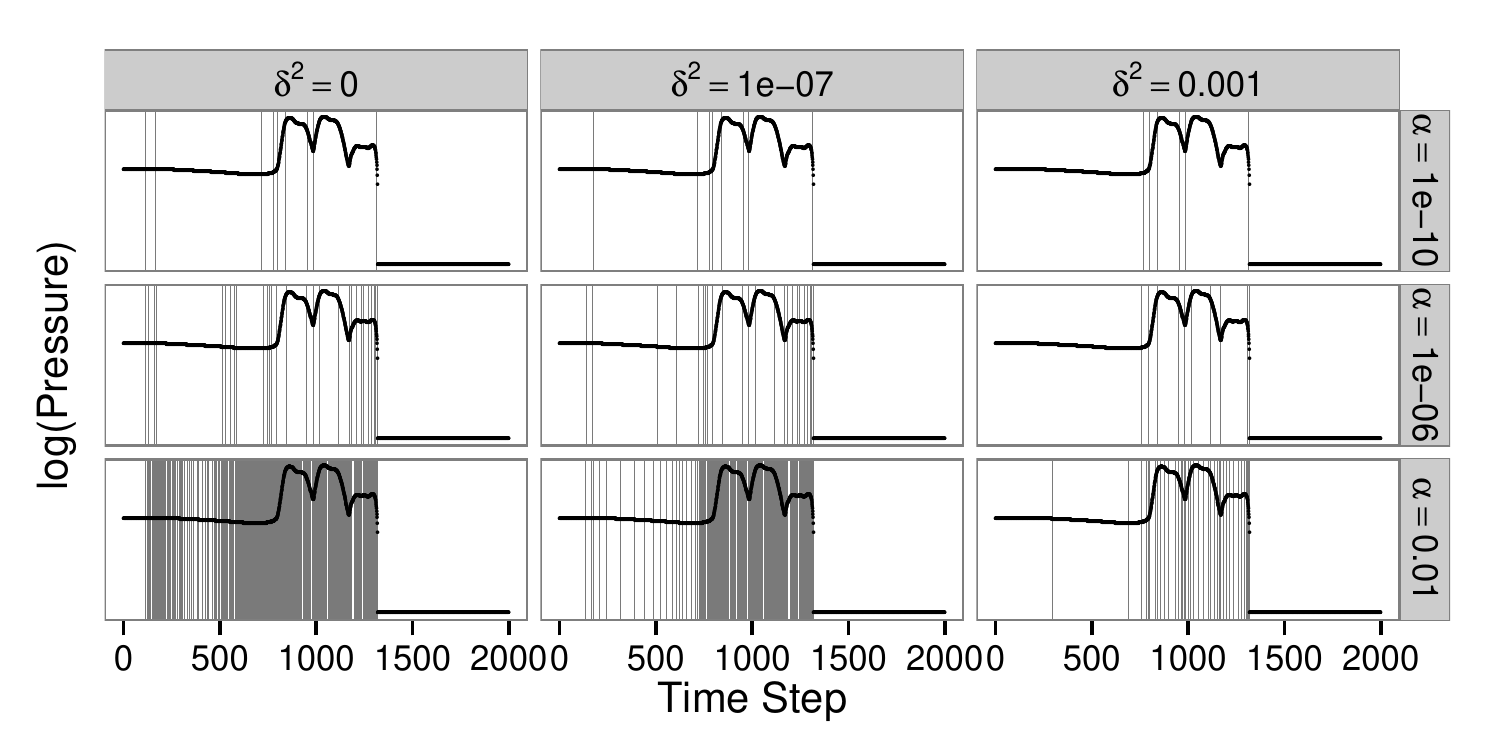}
\caption{Partitioning results for a grid of choices for the tuning parameters $\alpha$ and $\delta^2$. }
\label{fig:parSweep}
\end{center}
\end{figure}

Focusing first on the left side of \fig{fig:heatmaps}, we can see that both $\alpha$ and $\delta^2$ have an impact on the number of partitions.  Specifically, as $\alpha$ increases from top to bottom of the table, the algorithm chooses more partitions --- that is, a larger $\alpha$ rejects $H_0$ more often, just as we expect in typical hypothesis test settings. Likewise the algorithm choose more partitions as $\delta^2$ decreases from right to left.  However, the decisions on where to place the additional partitions are different for the two parameters. We can see this more clearly in \fig{fig:parSweep}, which shows the partitions selected for 9 of the table cells from \fig{fig:heatmaps}: those for $\alpha \in \{10^{-2}, 10^{-6}, 10^{-10}\}$ and $\delta^2 \in \{0, 10^{-7}, 10^{-3}\}$.  

We see that the parameter $\delta^2$ primarily acts by allowing the algorithm to ignore flat regions.  Setting $\delta^2$ to 0, as in the left column of \fig{fig:parSweep}, is equivalent to using the standard test in \eqref{eq:Ftest}. We see that this tends to insert partitions within the first 500 time steps, where the marginal variance is negligible as discussed in \sect{sec:LCROSS}.  As noted earlier, this region does have a subtle slope change, but it is small compared with the activity between time steps 750 and 1300, and we argue that adding partitions here does not improve the overall fit enough to justify the added cost.

Changing $\delta^2$ from 0 to a positive value corresponds to adding white noise with variance $\delta^2$.  The effect is to wash out small changes in the flatter regions of the simulation.  In each row of Figure \ref{fig:parSweep}, as $\delta^2$ increases from left to right fewer partitions are selected in the flat region of the first 500 time steps.    

Thus, $\delta^2$ provides a global effect that determines what sort of behavior is considered important.  This parameter generally scales with the overall dynamic range of the simulation.  From the form of our modified $F$-statistic in \eqref{eq:Ftestmod}, we can see that if the entire simulation were multiplied by a factor $\lambda$, then the algorithm would return the same decisions if $\delta^2$ were multiplied by $\lambda^2$. If the scientist has some prior knowledge that there are periods of active and inactive behavior in the simulation and some knowledge of how quiet the inactive periods are, $\delta^2$ can be set to accommodate this, i.e., to ignore those inactive periods if desired. 

The parameter $\alpha$ has a more local effect as it judges the differences in the fits on either side of the potential partition between \curr{} and \buff{}.  When $\delta^2$ is small or zero as in the first two columns of Figure \ref{fig:parSweep}, we see more partitions placed in the flat regions.  As $\alpha$ increases (moving down a column in Figure \ref{fig:parSweep}), partitions tend to be added in areas around other partitions, rather than in new areas.  That is, while $\delta^2$ plays a role in deciding which areas of the simulation are important in some sense, $\alpha$ guides how much detail will be provided in those areas. Eventually, as $\alpha$ is increased for a fixed $\delta^2$, partitions will be added in previously unpartitioned areas, as in the lower right panel.

Figure \ref{fig:parSweep} also provides a good demonstration of the greedy online nature of the algorithm.  Consider the third column with $\delta^2= 0.001$.  When all 2000 time steps of the simulation are viewed as a whole, you might expect an algorithm to add a partition in the trough between the second and third plateaus around time step 1250.  Instead, we can see in the second row that this partition doesn't get added unless some other earlier partitions are also included.  This is a direct result of the fact the algorithm can't see very far ahead.  The partition in the trough doesn't improve the local greedy fit unless a partition is added at the end of the second plateau before the sharp drop.  The partition at the end of the second plateau doesn't improve the fit unless a partition is added at the beginning of the plateau just after time step 1000.

\begin{figure}
\begin{center}
$\alpha = 10^{-8}$, $\delta^2 = 10^{-4}$: 7 partitions, $RSS_{\sf total} = 1207.75$ \\
\includegraphics[width=6in]{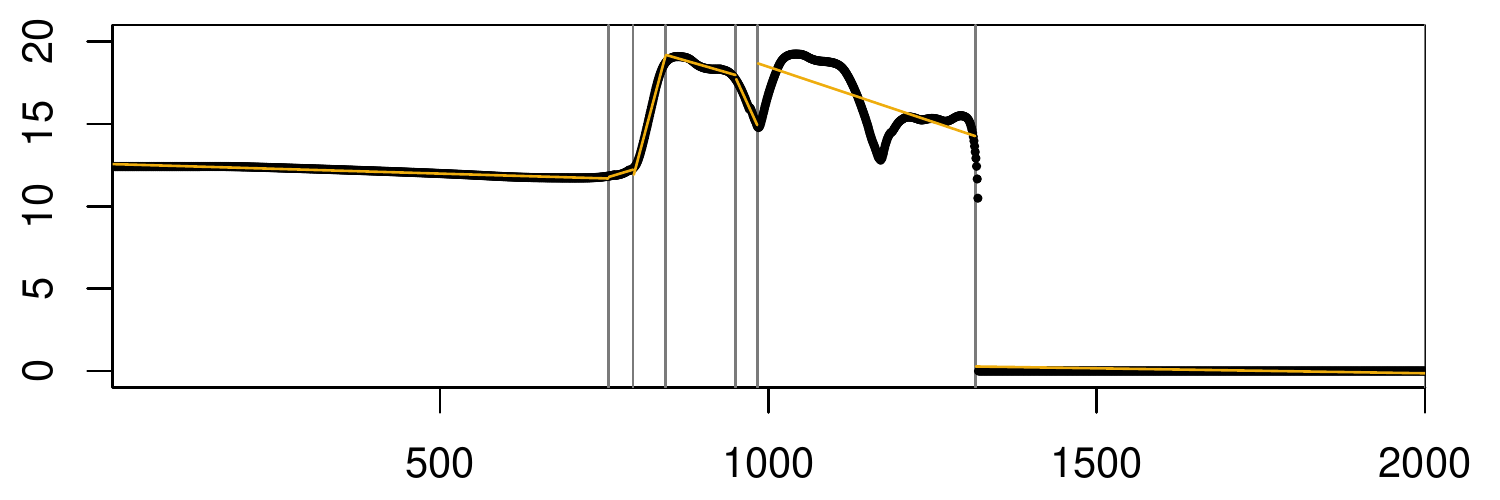}\\
$\alpha = 10^{-7}$, $\delta^2 = 10^{-4}$: 11 partitions, $RSS_{\sf total} = 38.33$\\
\includegraphics[width=6in]{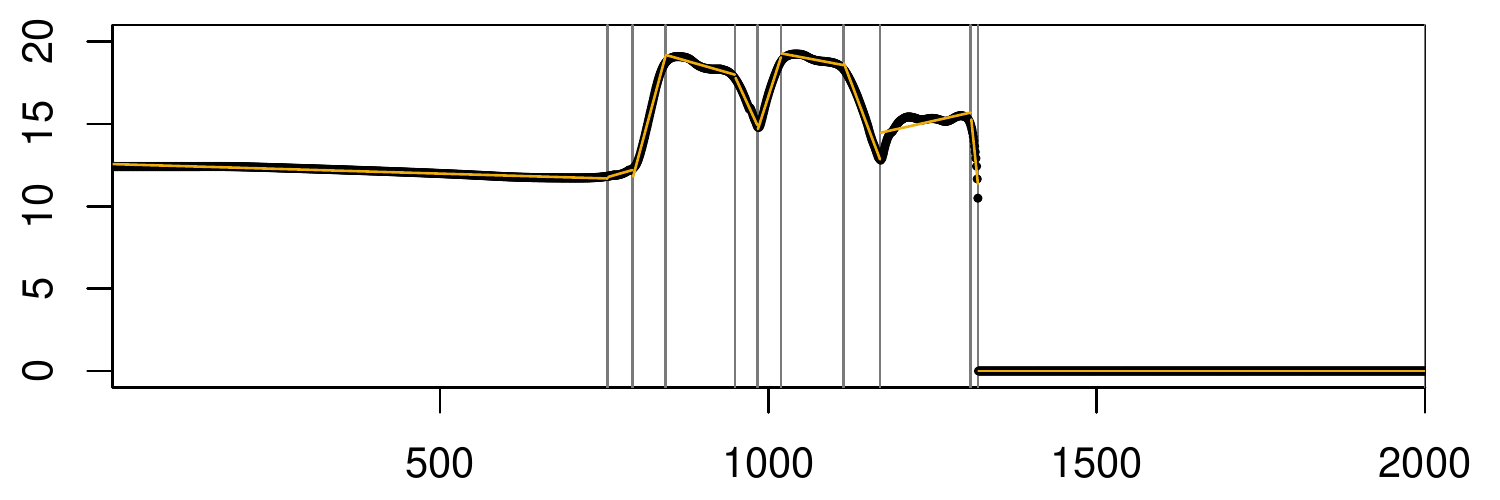}
\caption{In the top panel, the final partition chosen by our greedy online method starts a few time steps before the simulation flatlines at 0, resulting in a very high RSS for the final partition that accounts for almost half of the $RSS_{\sf total}$. In the bottom panel, with a larger $\alpha$ setting (and therefore more partitions selected), the final partition only contains time steps where the pressure variable has value 0, improving the overall fit substantially.}
\label{fig:badpick}
\end{center}
\end{figure}

We can see another impact of the greedy nature of our algorithm in the right panel of \fig{fig:heatmaps}, which reports the total RSS for the partitions chosen at each setting of $\alpha$ and $\delta^2$. There we see some abrupt changes, such as when $\alpha$ increases from $10^{-8}$ to $10^{-7}$ (third and fourth rows of the table) and produces a decrease of two orders of magnitude in the total RSS for many choices of $\delta^2$. This phenomenon is in part a result of the issue we first saw in \sect{algorithm}, namely that our greedy algorithm treats all the time steps in \buff{} as a single unit, which can cause it to miss the ``correct'' breakpoint by a few time steps. 
\fig{fig:badpick} shows the partitions chosen for these two $\alpha$ values with $\delta^2 = 10^{-4}$. We can see that in the top panel, the final partition chosen by the algorithm includes a few of the time steps that precede the drop to 0 pressure. This leads to an RSS of 553.92 for that final partition, nearly half the total RSS. (The majority of the remainder of the total RSS in this case is due to the poor fit in the next-to-last partition.) When $\alpha$ is increased in the bottom panel, leading to more partitions, the final partition contains only time steps where the pressure value is 0 and so fits the simulation perfectly with an RSS of 0. Similarly, when the RSS increases by an order of magnitude as $\alpha$ increases from 0.01 to 0.1 for many choices of $\delta^2$ in \fig{fig:heatmaps}, it's largely due to missing that final partition by a few time steps. As mentioned earlier, the fact that the RSS for a partition is unusually high can cue the scientists to check that part of the simulation.

Our discussion thus far has described the behavior or impact of the tuning parameters $\alpha$ and $\delta^2$ but not how to set them. Because the tuning parameters influence the tradeoff between the number of partitions required and the fidelity of the piecewise approximation, it's tempting to think about using a criterion such as the Akaike Information Criterion (AIC) to choose the setting. AIC is appealing since it captures the tradeoff between the number of parameters and the likelihood function. However, the same challenges in these deterministic computer simulations that led us to modify our $F$-statistic in \sect{sec:modF} make it difficult to calculate the AIC in the standard way. We will explore this more in future work, and in  \sect{sec:discussion} we briefly discuss other theoretical frameworks that could be useful for studying the algorithm and guiding parameter settings.

In the meantime we recommend that scientists invest a few runs of the simulation at different settings of $\alpha$ and $\delta^2$ to build a small version of the tables in \fig{fig:heatmaps} or to tune the parameters sequentially. 
An initial pass for a particular $\alpha$ and $\delta^2$ will return a set of partitions and the RSS within each partition.  If there are partitions with low error on both sides, while other partitions have large error, increasing $\delta^2$ will focus the partitions in the more active regions.  If more or fewer partitions are desired, increasing or decreasing $\alpha$ will help.  If this algorithm is being used as part of large suite of simulations which are expected to have broadly similar behavior, then the parameters can be tuned by rerunning a few of the simulations and then fixing their settings for the remaining runs.  This would limit the increase in total computation while making substantially more efficient use of computational resources over the entire suite. 

\section{Discussion} \label{sec:discussion}
We have demonstrated a simple, computationally efficient statistical approach that can be embedded in a complex computer simulation to identify important time steps, dramatically reduce data storage requirements, and facilitate later reconstruction and analysis of the simulation. We developed this approach in anticipation of the computational demands of exascale simulations. 

The details of the hypothesis test raise a number of important issues that should be explored, such as the violations of the typical assumptions for hypothesis testing.  We addressed this issue here by introducing the tuning parameter $\delta^2$ that inflates the variance of the regression model. We plan to explore ways to optimize the choice of $\alpha$ and $\delta^2$, such as by choosing the values that best approximate the hypothesis test that would result from assuming a correlated error structure (e.g., a Gaussian process) with some assumed correlation parameters. We might also explore whether multiple hypothesis testing presents a problem in our context since we are performing our hypothesis test repeatedly at each simulation time step. This is likely less of a problem here than in typical settings because of the deterministic nature of the error in these simulations.  At this point, we choose to treat our test statistic as a useful measure for detecting practical changes in these large-scale simulations, rather than assigning the usual interpretation of statistical significance.

To provide a theoretical framework for studying our algorithm and suggesting settings for our tuning parameters $\alpha$ and $\delta^2$, we can consider models that produce smooth realizations like those seen in our examples and study how our algorithm performs on them as a function of the autocorrelation.  Gaussian processes are often used to represent simulation output and would be useful in this context as well.  Process convolution models could be used to produce nonstationary sequences that would help us understand more precisely how our method identifies partitions.


An obvious extension to consider is the incorporation of slightly more complicated models than the piecewise linear fits of scalars presented here. For instance, piecewise polynomial models would likely improve the approximation, though they would also increase the computational burden. Another interesting avenue is to relax our assumption of independence between scalars and allow any correlation to be modeled as well.

Finally, we note that our method identifies important time steps in a scientifically agnostic manner --- we are not using any domain-specific knowledge to identify important features in the simulation.  As statisticians at a collaborative science laboratory, we consider our approach to be a useful empirical tool for exploration intended to spur incorporation of scientific knowledge into the analysis and subsequent model development. When possible, this approach should be augmented with domain knowledge to move beyond a simple empirical assessment to produce a science-driven result. The ideal scenario would include input from the scientists regarding expected features and behavior of interest, but domain scientists may not always know how the physics and scales in their simulation interact.  Further, they may not have thoroughly tested their codes over complete ranges of their input parameters, so bugs may remain.  It is for these situations, and to provide a general tool for many domains, that we have developed the current method.  As the need for \insitu\/ analysis continues to grow, we anticipate that future work will develop methods that are tailored to specific scientific questions.

\bibliographystyle{apalike}
\bibliography{LinearFits}

\end{document}